# Search for direct CP violation in the decays $K^{\pm} \to 3\pi^{\pm}$

The NA48/2 Collaboration

J.R. Batley, A.J. Culling, G. Kalmus, C. Lazzeroni, D.J. Munday, M.W. Slater, S.A. Wotton

*Cavendish Laboratory, University of Cambridge, Cambridge, CB3 0HE, U.K.* [1]

R. Arcidiacono, G. Bocquet, N. Cabibbo, A. Ceccucci, D. Cundy [2], V. Falaleev, M. Fidecaro, L. Gatignon, A. Gonidec, W. Kubischta, A. Norton, A. Maier, M. Patel, A. Peters

*CERN, CH-1211 Genève 23, Switzerland*

S. Balev, P.L. Frabetti, E. Goudzovski, P. Hristov [3], V. Kekelidze [3], V. Kozhuharov, L. Litov, D. Madigozhin, E. Marinova, N. Molokanova, I. Polenkevich, Yu. Potrebenikov, S. Stoynev, A. Zinchenko

*Joint Institute for Nuclear Research, Dubna, Russian Federation*

E. Monnier [4], E. Swallow, R. Winston

*The Enrico Fermi Institute, The University of Chicago, Chicago, Illinois, 60126, U.S.A.*

P. Rubin, A. Walker

*Department of Physics and Astronomy, University of Edinburgh, JCMB King's Buildings, Mayfield Road, Edinburgh, EH9 3JZ, U.K.*

W. Baldini, A. Cotta Ramusino, P. Dalpiaz, C. Damiani, M. Fiorini, A. Gianoli, M. Martini, F. Petrucci, M. Savrié, M. Scarpa, H. Wahl

*Dipartimento di Fisica dell'Università e Sezione dell'INFN di Ferrara, I-44100 Ferrara, Italy*

A. Bizzeti [5], M. Calvetti, E. Celeghini, E. Iacopini, M. Lenti, F. Martelli [6], G. Ruggiero [3], M. Veltri [6]

*Dipartimento di Fisica dell'Università e Sezione dell'INFN di Firenze, I-50125 Firenze, Italy*

M. Behler, K. Eppard, K. Kleinknecht, P. Marouelli, L. Masetti, U. Moosbrugger, C. Morales Morales, B. Renk, M. Wache, R. Wanke, A. Winhart

*Institut für Physik, Universität Mainz, D-55099 Mainz, Germany* [7]

D. Coward [8], A. Dabrowski, T. Fonseca Martin [3], M. Shieh, M. Szleper, M. Velasco, M.D. Wood [9]

*Department of Physics and Astronomy, Northwestern University, Evanston Illinois 60208-3112, U.S.A.*

G. Anzivino, P. Cenci, E. Imbergamo, A. Nappi, M. Pepe, M.C. Petrucci, M. Piccini, M. Raggi, M. Valdata-Nappi

*Dipartimento di Fisica dell'Università e Sezione dell'INFN di Perugia, I-06100 Perugia, Italy*

C. Cerri, G. Collazuol, F. Costantini, L. DiLella, N. Doble, R. Fantechi, L. Fiorini, S. Giudici, G. Lamanna, I. Mannelli, A. Michetti, G. Pierazzini, M. Sozzi

*Dipartimento di Fisica dell'Università, Scuola Normale Superiore e Sezione dell'INFN di Pisa, I-56100 Pisa, Italy*

B. Bloch-Devaux, C. Cheshkov [3], J.B. Chèze, M. De Beer, J. Derré, G. Marel, E. Mazzucato, B. Peyaud, B. Vallage

*DSM/DAPNIA - CEA Saclay, F-91191 Gif-sur-Yvette, France*

M. Holder, M. Ziolkowski
*Fachbereich Physik, Universität Siegen, D-57068 Siegen, Germany* [10]

S. Bifani, C. Biino, N. Cartiglia, M. Clemencic [3], S. Goy Lopez, F. Marchetto
*Dipartimento di Fisica Sperimentale dell'Università e Sezione dell'INFN di Torino, I-10125 Torino, Italy*

H. Dibon, M. Jeitler, M. Markytan, I. Mikulec, G. Neuhofer, L. Widhalm
*Österreichische Akademie der Wissenschaften, Institut für Hochenergiephysik, A-10560 Wien, Austria* [11]

**Abstract**

We report a measurement of the direct CP violation asymmetry parameter $A_g$ in charged kaon decays $K^\pm \to \pi^\pm\pi^+\pi^-$ by the NA48/2 experiment at the CERN SPS. The experiment has been designed not to be limited by systematics in the asymmetry measurement. Using $1.67 \times 10^9$ such decays collected during the 2003 run, the charge asymmetry in the Dalitz plot linear slope parameter $g$ has been measured to be $A_g = (1.7 \pm 2.9) \times 10^{-4}$. The precision of the result is limited by the statistics used.

[1]Funded by the U.K. Particle Physics and Astronomy Research Council

[2]Present address: Istituto di Cosmogeofisica del CNR di Torino, I-10133 Torino, Italy

[3]Present address: CERN, CH-1211 Genève 23, Switzerland

[4]Also at Centre de Physique des Particules de Marseille, IN2P3-CNRS, Université de la Méditerranée, Marseille, France

[5]Also Istituto di Fisica, Università di Modena, I-41100 Modena, Italy

[6]Istituto di Fisica, Università di Urbino, I-61029 Urbino, Italy

[7]Funded by the German Federal Minister for Education and research under contract 05HK1UM1/1

[8]Permanent address: SLAC, Stanford University, Menlo Park, CA 94025, U.S.A.

[9]Present address: UCLA, Los Angeles, CA 90024, U.S.A.

[10]Funded by the German Federal Minister for Research and Technology (BMBF) under contract 056SI74

[11]Funded by the Austrian Ministry for Traffic and Research under the contract GZ 616.360/2-IV GZ 616.363/2-VIII, and by the Fonds für Wissenschaft und Forschung FWF Nr. P08929-PHY

# 1 Introduction

More than 40 years after its discovery [1], the phenomenon of CP violation still plays a central role in present and future particle physics investigations. For a long time the effect seemed to be confined to a peculiar sector of particle physics. However, two important discoveries took place recently. In the late 1990's the existence of direct CP violation was firmly established: following an earlier indication by NA31 [2], the NA48 and KTeV experiments demonstrated the existence of direct CP violation with high significance [3, 4]. This was done by measuring a non-zero $\varepsilon'/\varepsilon$ parameter in the decays of neutral kaons into two pions. In 2001, the $B$-factory experiments Babar and Belle measured CP violation in the system of neutral $B$ mesons [5] and in 2004 also direct CP violation in $B$ decays has been found [6].

In order to explore possible non-Standard Model (SM) enhancements to heavy-quark loops, which are at the core of direct CP-violating processes, all manifestations of direct CP violation have to be studied experimentally. In kaons, besides the parameter $\varepsilon'/\varepsilon$ already measured in $K_{L,S} \to 2\pi$ decays, the most promising complementary observables are decay rates of GIM-suppressed kaon decays, which proceed through flavour-changing neutral currents, and the asymmetry between $K^+$ and $K^-$ decays into three pions.

Direct CP violation manifesting itself as an asymmetry in two CP-conjugate decay amplitudes (as in the $K^\pm \to 3\pi$ case) is important as a strong qualitative test of the way the SM accommodates CP violation. However, its quantitative exploration to constrain the fundamental parameters of the theory is difficult due to non-perturbative hadronic effects. Still, an intense theoretical programme is under way to improve such predictions, which will ultimately allow the direct CP violation measurements to be used as strong quantitative constraints on the SM.

The $K^\pm \to 3\pi$ matrix element squared is usually parameterized by a polynomial expansion [7]:

$$|M(u,v)|^2 \sim 1 + gu + hu^2 + kv^2, \tag{1}$$

where $g$, $h$, $k$ are the so called linear and quadratic Dalitz plot slope parameters ($|h|, |k| \ll |g|$) and the two Lorentz invariant kinematic variables $u$ and $v$ are defined as

$$u = \frac{s_3 - s_0}{m_\pi^2}, \quad v = \frac{s_2 - s_1}{m_\pi^2}, \quad s_i = (P_K - P_i)^2, \; i = 1,2,3; \quad s_0 = \frac{s_1 + s_2 + s_3}{3}. \tag{2}$$

Here $m_\pi$ is the pion mass, $P_K$ and $P_i$ are the kaon and pion four-momenta, the indices $i = 1, 2$ correspond to the two identical ("even") pions and the index $i = 3$ to the pion of different charge (the "odd" pion). A term linear in $v$ is forbidden in (1) due to symmetry considerations. A difference of slope parameters $g^+$ and $g^-$ describing positive and negative kaon decays, respectively, is a manifestation of direct CP violation usually defined by the corresponding slope asymmetry

$$A_g = (g^+ - g^-)/(g^+ + g^-) \approx \Delta g/(2g), \tag{3}$$

where $\Delta g$ is the slope difference and $g$ is the average slope. SM predictions for $A_g$ vary from a few $10^{-6}$ to a few $10^{-5}$ [8]. The asymmetry of integrated decay rates is expected to be strongly suppressed with respect to the slope asymmetry [9]. Existing theoretical calculations involving processes beyond the SM [10] allow a wider range of $A_g$, including substantial enhancements up to a few $10^{-4}$. Several experiments have searched for the asymmetry $A_g$ in both $\pi^\pm\pi^+\pi^-$ and $\pi^\pm\pi^0\pi^0$ kaon decay modes [11] and the upper limits reached so far are at the level of a few $10^{-3}$.

The NA48/2 experiment is carried out in the framework of kaon physics programme at the CERN SPS. Its primary aim is to measure the asymmetries $A_g$ and $A_g^0$ in $K^\pm \to \pi^\pm\pi^+\pi^-$ and $K^\pm \to \pi^\pm\pi^0\pi^0$ decays, respectively, with a precision at least one order of magnitude better than the existing limits. A preliminary result of the NA48/2 $A_g$ measurement based on 2003 data has been reported in [12]. The present result based on improved analysis and slightly larger statistics supersedes this in precision.

# 2 Beams and detectors

High precision measurement of $A_g$ (at the level of $10^{-4}$) requires not only high statistics, but also a dedicated experimental approach. A beam line transporting two simultaneous charged beams of opposite signs was designed and built as a key element leading to cancellations of main systematic uncertainties, allowing decays of $K^+$ and $K^-$ to be recorded at the same time. Regular alternation of magnetic fields in all the beam line elements and the spectrometer magnet was adopted to symmetrize the acceptance for the two beams. The layout of the beams and detectors is shown schematically in Figure 1.

The setup is described in a right-handed orthogonal coordinate system with the $z$ axis directed downstream along the beam, and the $y$ axis directed vertically up.

The beams are produced by 400 GeV protons impinging on a beryllium target of 40 cm length and 2 mm diameter at zero incidence angle. Charged particles with momentum $(60 \pm 3)$ GeV/$c$ are selected symmetrically for positive and negative particles by an achromatic system of four dipole magnets with null total deflection, which splits the two beams in the vertical plane and then recombines them on a common axis. They then pass through a defining collimator and a series of four quadrupoles designed to produce horizontal and vertical charge-symmetric focusing of the beams towards the detector. Finally they are again split and recombined in a second achromat, where three stations of a MICROMEGAS type detector operating in TPC mode form the kaon beam spectrometer (KABES) [13] (however, not used in the present analysis). With $7 \times 10^{11}$ protons per burst of $\sim 4.8$ s duration incident on the target, the positive (negative) beam flux at the entrance of the decay volume is $3.8 \times 10^7$ $(2.6 \times 10^7)$ particles per pulse, of which 5.7% (4.9%) are $K^+$ $(K^-)$. The $K^+/K^-$ flux ratio is about 1.8.

Downstream of the second achromat both beams follow the same path. After passing the cleaning and the final collimators they enter the decay volume, housed in a 114 m long vacuum tank with a diameter of 1.92 m for the first 65 m, and 2.4 m for the rest. The fraction of beam kaons decaying in this volume is about 22%. The beams are steered to be collinear to within about 1 mm throughout the entire decay volume. Superposition of the beams symmetrizes the detector acceptances and contributes to the reduction of systematic biases. Focusing by the system of quadrupoles aims to obtain a minimal transverse size in the region of the detector, which minimizes sensitivity to any transverse structure of the beams.

The decay volume is followed by a magnetic spectrometer [14] used for the reconstruction of $K^\pm \to 3\pi^\pm$ decays. The spectrometer is housed in a tank filled with helium at nearly atmospheric pressure, separated from the vacuum tank by a thin (0.0031 radiation lengths) *Kevlar*-composite window. A thin-walled aluminium beam tube of $\sim 16$ cm diameter traversing the centre of the detector allows the undecayed beam particles and the muon halo from decays of beam pions to continue their path in vacuum. The magnetic spectrometer consists of four drift chambers (DCH): two located upstream and two downstream of the dipole magnet, which provides a horizontal transverse momentum kick of about 120 MeV/$c$ to the charged particles. The DCHs have an octagonal shape with transverse size of about 2.8 m and fiducial area of about 4.5 m$^2$. Each chamber is composed of eight planes of sense wires arranged in four couples of staggered planes (so called views) oriented horizontally, vertically, and along each of the two orthogonal 45° directions. The momentum resolution of the magnetic spectrometer is $\sigma_p/p = 1.02\% \oplus 0.044\% p$ ($p$ is expressed in GeV/$c$), corresponding to a reconstructed $3\pi$ invariant mass resolution of about 1.7 MeV/$c^2$. The magnetic spectrometer is followed by a scintillator hodoscope consisting of a plane of horizontal and a plane of vertical strips, each plane is arranged in four quadrants (strip widths range from 6.5 cm for central counters to 9.9 cm for peripheral ones). The hodoscope is in turn followed by a liquid krypton electromagnetic calorimeter (LKr), a hadronic calorimeter (HAC), and a muon detector (MUV).

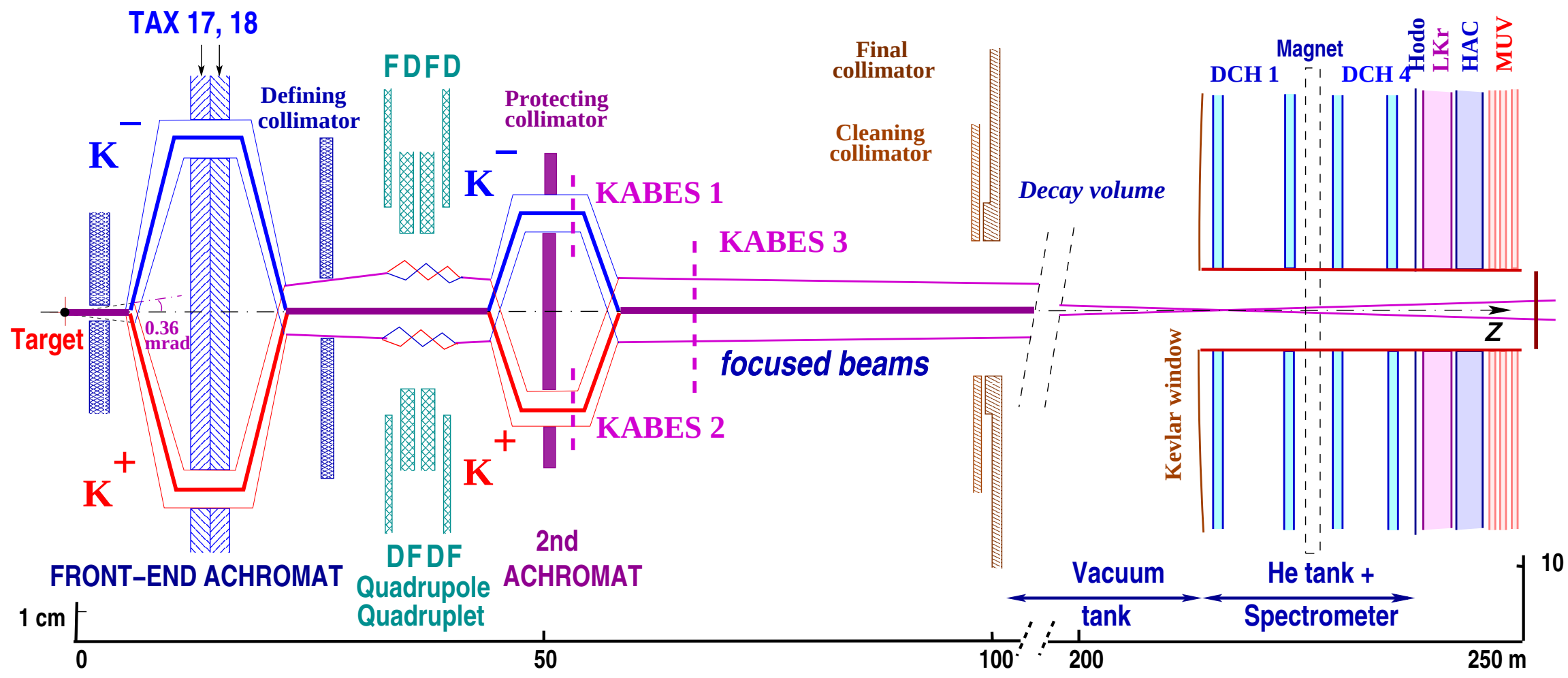

Figure 1: Schematic lateral view of the NA48/2 beam line (TAX17,18: motorized beam dump/collimators used to select the momentum of the $K^+$ and $K^-$ beams; DFDF: focusing quadrupoles, KABES1–3: kaon beam spectrometer stations), decay volume and detector (DCH1–4: drift chambers, Hodo: hodoscope, LKr: EM calorimeter, HAC: hadron calorimeter, MUV: muon veto). The vertical scale changes for the two parts of the figure.

The $K^\pm \to 3\pi^\pm$ decays are triggered with a two-level system. At the first level (L1), the rate of $\sim$ 500 kHz is reduced to $\sim$ 100 kHz by requiring coincidences of hits in the two planes of the scintillator hodoscope in at least two quadrants. The second level (L2) is based on a hardware system computing coordinates of DCH hits from DCH drift times and a farm of asynchronous microprocessors performing fast reconstruction of tracks and running the decision taking algorithm. The L2 algorithm requires at least two tracks to originate in the decay volume with the closest distance of approach of less than 5 cm. L1 triggers not satisfying this condition are examined further and accepted if there is a reconstructed track which is not kinematically compatible with a $\pi^\pm\pi^0$ decay of a $K^\pm$ having momentum of 60 GeV/$c$ directed along the beam axis. The resulting trigger rate is about 10 kHz.

The description of other components of the NA48 detector less relevant for the present

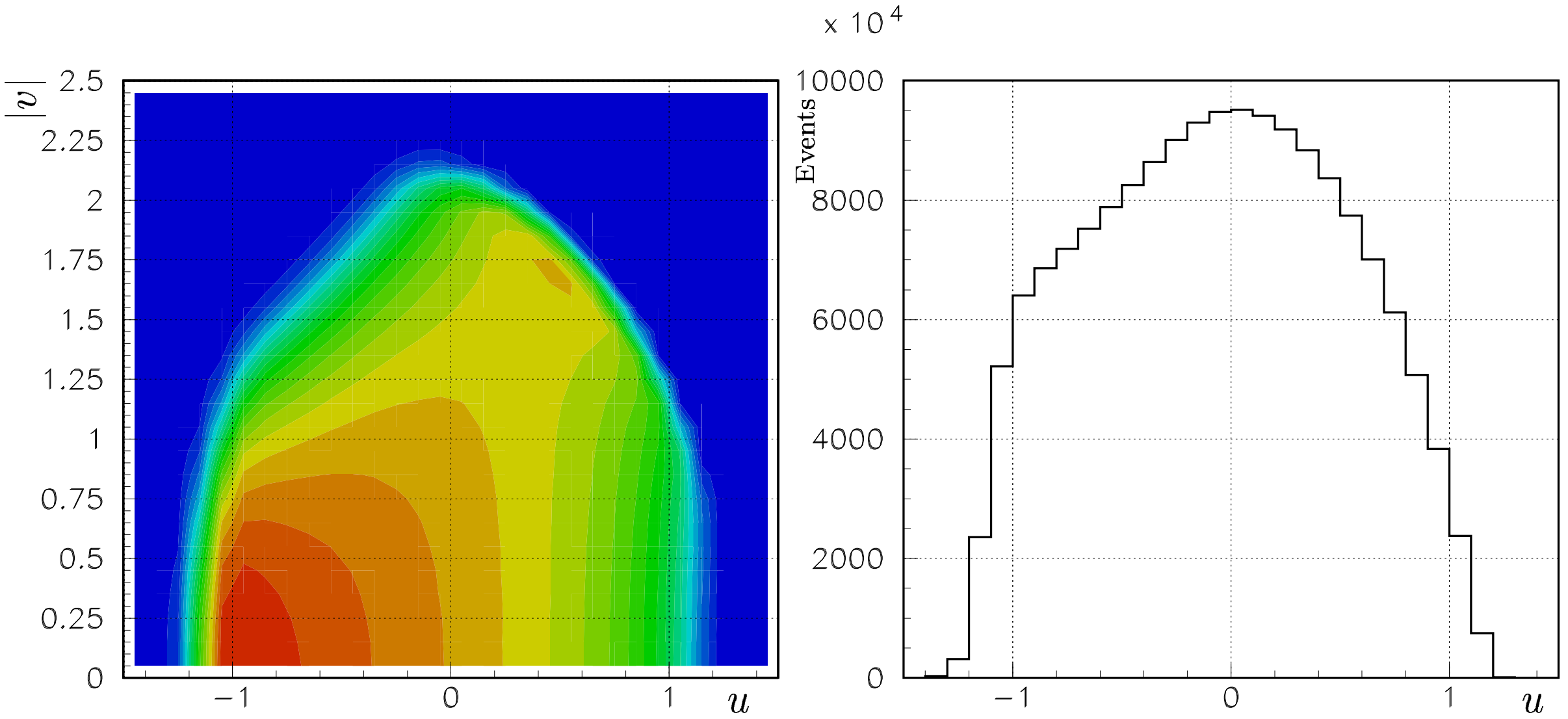

Figure 2: Left: Normalized distribution of the selected events in the kinematical variables $(u,|v|)$, right: $u$ spectrum of the selected events.

analysis can be found elsewhere [3].

NA48/2 collected data during two runs in 2003 and 2004, with $\sim$50 days of efficient data taking each. About $7 \times 10^9$ triggers were recorded during the 2003 run, which provides the data sample described in this paper.

# 3 Measurement method

A distribution of the finally selected $K_{3\pi}$ events[1] in the kinematical variables $(u,|v|)$ and its $u$ projection are presented in Fig. 2. The resolution on $u$ varies over the Dalitz plot, but never exceeds 0.06, which is less than the chosen bin width of 0.10. The measurement method is based on comparing the $u$ spectra of $K^+$ and $K^-$ decays $N^+(u)$ and $N^-(u)$. Given the actual values of the slope parameters $g$, $h$, and $k$ [7] and the pecision of the current measurement, the ratio of $u$ spectra of $R(u) = N^+(u)/N^-(u)$ is in good approximation proportional to $(1 + \Delta g \cdot u)$, so $\Delta g$ can be extracted from a linear fit to the ratio $R(u)$, and $A_g = \Delta g/2g$ can be evaluated.

Charge symmetrization of the experimental conditions is to a large extent achieved by using simultaneous and collinear $K^+$ and $K^-$ beams with similar momentum spectra. However, the presence of magnetic fields in both the beam line (achromats, focusing quadrupoles, etc.) and the magnetic spectrometer, combined with some asymmetries in detector performance, introduces residual charge asymmetries. In order to equalize the local effects on the acceptance, the polarities of all the magnets in the beam line were reversed during the data taking on an approximately weekly basis (corresponding to the periodicity of SPS technical stops), and the polarity of the spectrometer magnet was reversed approximately once per day. Data collected over a period with all the four possible setup configurations (i.e. combinations of beam line and spectrometer magnet polarities) spanning approximately two weeks of efficient data taking represent a "supersample", which is treated as an independent and self-consistent set of data. During the 2003 run, four supersamples (numbered 0 to 3) were collected.

[1]The essential selection criteria will be discussed in the next chapter.

Each supersample contains four sets of simultaneously collected $K^+ \to 3\pi$ and $K^- \to 3\pi$ samples corresponding to the four different setup configurations (totally eight data samples). To measure the charge asymmetry, exploiting the cancellations of systematic biases emerging due to polarity reversals, the following "quadruple ratio" involving the eight corresponding $u$ spectra, composed as a product of four $R(u) = N^+(u)/N^-(u)$ ratios with opposite kaon sign, and deliberately chosen setup configurations in numerator and denominator, is considered:

$$R_4(u) = R_{US}(u) \cdot R_{UJ}(u) \cdot R_{DS}(u) \cdot R_{DJ}(u). \quad (4)$$

In the notation for the $R(u)$ ratios, the indices $U$ ($D$) correspond to beam line polarities corresponding to $K^+$ passing along the upper (lower) path in the achromats, respectively, while the indices $S$ ($J$) represent spectrometer magnet polarities (opposite for $K^+$ and $K^-$) corresponding to the "even" (i.e. the two identical) pions being deflected to negative (positive) $x$, i.e. towards the Salève (Jura) mountains, respectively. A fit of the quadruple ratio (4) with a linear function $f(u) = n \cdot (1 + 4\Delta g \cdot u)$ results in two parameters: the normalization $n$ and the difference of slopes $\Delta g$. The normalization is sensitive to the $K^+/K^-$ flux ratio, while $\Delta g$ is not.

The quadruple ratio technique logically completes the procedure of magnet polarity reversal, and allows a three-fold cancellation of systematic biases:

- due to spectrometer magnet polarity reversal, local detector biases cancel between $K^+$ and $K^-$ samples with decay products reaching the same parts of the detector in each of the four ratios $R(u)$ appearing in the quadruple ratio $R_4(u)$;
- due to the simultaneous beams, global time-variable biases cancel between $K^+$ and $K^-$ samples in the product of $R_S(u)$ and $R_J(u)$ ratios;
- due to beam line polarity reversal, local beam line biases, resulting in slight differences in beam shapes and momentum spectra, largely cancel between the $R_U(u)$ and $R_D(u)$ ratios.

Remaining systematic biases due to the presence of permanent magnetic fields (Earth's field, vacuum tank magnetization) are minimized by maintaining azimuthal symmetry of the geometrical acceptance by appropriate cuts.

The method is independent of the $K^+/K^-$ flux ratio and the relative sizes of the samples collected with different magnet configurations. However, the statistical precision is limited by the smallest of the samples involved, so the balance of sample sizes was controlled during the data taking. The result remains sensitive only to time variations of asymmetries in the experimental conditions which have a characteristic time smaller than corresponding field alternation period, and in principle should be free of systematic biases.

Due to the method described above, no Monte Carlo (MC) corrections to the acceptance are required. Nevertheless, a detailed GEANT-based [15] MC simulation was developed as a tool for systematic studies, including full detector geometry and material description, simulation of time-variable local DCH inefficiencies, time variations of the beam geometry and DCH alignment. A large MC production was carried out, providing a sample of a size comparable to that of the data.

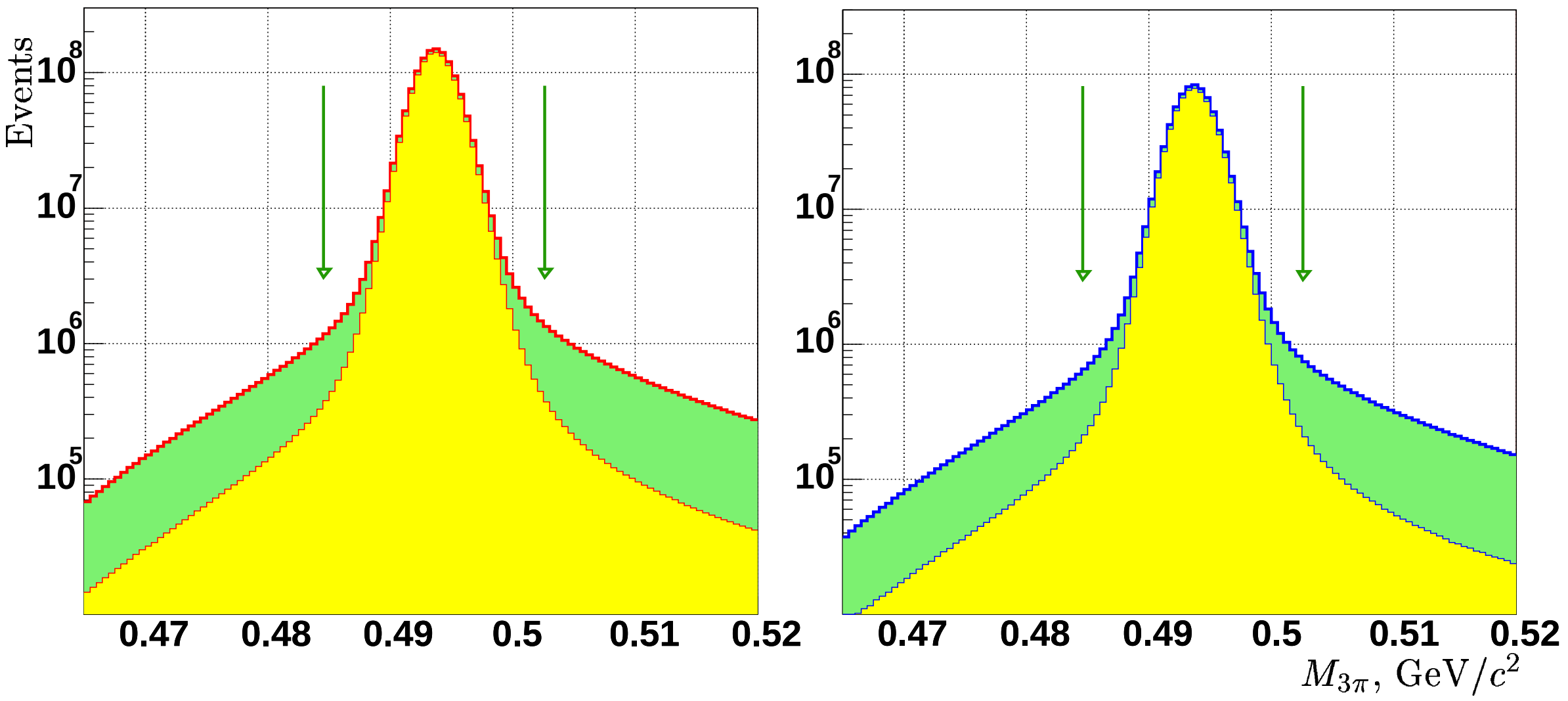

Figure 3: Spectra of reconstructed $3\pi^{\pm}$ masses corresponding to $K^+$ (left) and $K^-$(right) decays. Dark areas indicate events with a muon identified by muon veto system.

# 4 Data analysis

Several stages of data compaction and filtering were necessary in order to reduce the 2003 data volume from $\sim 100$ TB of raw data to 619 GB of the final sample, while reducing the number of events recorded to $2.43 \times 10^9$. The main requirement of the filtering algorithm is the presence of at least 3 reconstructed tracks, which cuts away 55% of the collected triggers.

Tracks are reconstructed from hits in DCHs using the measured magnetic field map rescaled according to the recorded current in the spectrometer analyzing magnet. Three-track vertices compatible with a $K \to 3\pi$ decay topology are reconstructed by extrapolation of track segments from the upstream part of the spectrometer back into the decay volume, taking into account the stray magnetic fields due to the Earth and magnetization of the vacuum tank. The kinematics of the events is calculated using measured momenta and track directions at the vertex. The stray field correction is based on a three-dimensional field map measured in the entire vacuum tank. It reduces the observed azimuthal variation of the reconstructed invariant mass by more than an order of magnitude to $\sim 100$ keV/$c^2$.

Event selection includes rejection of trigger buffer overflow[2] events, requirements on vertex charge, quality, and position (within the decay volume, and laterally within the beam), limits on kaon momentum: 54 GeV/$c$ $< P_K <$ 66 GeV/$c$ and reconstructed $3\pi$ invariant mass: $|M_{3\pi} - M_K| < 9$ MeV/$c^2$ (the latter is roughly five times the corresponding resolution). The selection leaves a sample of $1.87 \times 10^9$ $K_{3\pi}$ events which is practically background free, as $K_{3\pi}$ is the dominant three-track decay mode. The distribution of reconstructed $3\pi$ invariant mass is presened in Fig. 3.

*Fine alignment of the magnetic spectrometer*

The transverse positions of DCHs and individual wires were realigned every 2–4 weeks of data taking using data collected in special runs in which muon tracks were recorded with no magnetic field in the spectrometer. This allows an alignment precision of $\sim 30$ $\mu$m to be reached. However, time variations of DCH alignment on a shorter time scale can potentially bias the asymmetry, since an uncorrected shift of a DCH along the $x$ axis leads to charge-antisymmetric mismeasurement of the momenta. An unambiguous measure of the residual misalignment is the difference between the average reconstructed $3\pi$ invariant masses for $K^+$ and $K^-$ decays ($\Delta\overline{M}$). A 1 $\mu$m shift of a DCH along the $x$ axis typically induces a mass shift of $\Delta\overline{M} \sim 1$ keV$/c^2$ (the proportionality factor varies among the DCHs). Monitoring of $\Delta\overline{M}$ revealed significant (up to 200 $\mu$m) movements of the DCHs between individual alignment runs. Introduction of time-dependent corrections to the measured momenta based on the observed $\Delta\overline{M}$ reduces the effect on the slope difference by more than an order of magnitude to a negligible level of $\delta(\Delta g) < 0.1 \times 10^{-4}$.

[2]Trigger buffer overflow indicates that a corresponding event has an enhanced number of hits in DCHs and therefore saturates the L2 trigger processor leading to significant reduction of the efficiency.

*Correction for beam geometry instabilities*

The most important feature determining the geometric acceptance is the beam tube traversing the centre of all DCHs. Moreover, the beam optics can control the average transverse beam positions to $\pm 1$ mm, leaving a sizable random charge-asymmetric bias to the acceptance. In order to compensate for this effect, inner radius cuts $R > 11.5$ cm are introduced for pion impact points at the first and the last DCHs with respect to the actual average measured beam position in DCH1 and its extrapolation from upstream of the magnet to DCH4. These cuts cost 12% of the statistics, leading to a final sample of $1.67 \times 10^9$ events. The minimum radius value of 11.5 cm is chosen to ensure that the beam tube and the adjacent central insensitive areas of the DCHs are securely excluded by the cut. The average beam positions are continuously monitored separately for $K^+$ and $K^-$ by calculating the momentum-weighted centre of gravity of the three pions in the planes of DCHs upstream of the magnet. A bias introduced by the fact that the average beam positions are themselves affected by the acceptance is negligible. In addition to the time variation of the average positions, also the dependencies of the beam position on kaon momentum ($\sim 1$ mm in the horizontal plane, $\sim 1$ cm in the vertical plane) and time in spill ($\sim 1$ mm) are monitored and taken into account. A precision of 100 $\mu$m in the determination of beam position is sufficient to reduce systematic effects to a negligible level.

Residual charge-asymmetric effects originate from permanent (irreversible) magnetic fields in the decay volume (vacuum tank magnetization, Earth field) coupling to time-dependent DCH inefficiencies and beam migrations. The corresponding fake asymmetries have been estimated not to exceed $\delta(\Delta g) = 0.3 \times 10^{-4}$.

*Trigger efficiency correction*

Only charge-asymmetric trigger inefficiencies dependent on $u$ can possibly bias the measurement. Inefficiencies of the trigger components are measured as functions of $u$ using control data samples from low bias triggers collected along with the main triggers. Control trigger condition for L1 efficiency measurement requires at least one coincidence of hits in the two planes of the scintillator hodoscope. Control triggers for L2 efficiency measurement are L1 triggers recorded regardless of L2 reply. The statistics of each of the two control samples is roughly 1% of the main sample. Thus time variations of inefficiencies can be accounted for and the statistical errors on the inefficiency measurements are propagated into the final result.

The inefficiency of the L1 trigger, due to hodoscope inefficiency, was measured to be $0.9 \times 10^{-3}$ and found to be stable in time. Due to temporary malfunctioning of a single hodoscope channel, a part of supersample 3 data has an inefficiency of $3.2 \times 10^{-3}$, which was reduced to the common level in the selected data sample by applying appropriate geometric cuts (relative to beam position) on the hodoscope surface for the whole supersample (thus preserving the acceptance symmetry), at the cost of 2.5% of the total statistics. Due to the time stability of the inefficiency, no correction is applied, and an uncertainty of $\delta(\Delta g) = 0.4 \times 10^{-4}$, limited by the statistics of the control sample, is attributed.

For the L2 trigger, corrections to $u$ spectra are introduced for the rate-independent part of the inefficiency, which is due to local DCH inefficiencies, which affect the trigger more than the reconstruction due to lower redundancy. This part of the inefficiency varied between 0.4% and 1.5% at the beginning of the run and then became stable at the level of 0.06% after the final version of the L2 algorithm was reached. The trigger buffer overflow cut was crucial to reduce the inefficiency. The L2 inefficiency correction to the whole statistics amounts to $\delta(\Delta g) = (-0.5 \pm 0.5) \times 10^{-4}$, where the error is statistical due to the limited size of the control sample. The symmetry of the rate-dependent part of the inefficiency of $\sim 0.2\%$ was checked separately with MC simulation of pile-up effects and a study of the dependence of the result on the number of allowed accidental tracks.

*Asymmetry fits and cross-checks*

After applying the corrections described above, $\Delta g$ is extracted by fitting the quadruple ratio of the $u$ spectra (4) independently for each supersample. The corresponfing quadruple ratio fits are presented in Figure 4. The numbers of events selected in each supersample, the "raw" values of $\Delta g$ obtained without applying the trigger corrections and the final $\Delta g$ with the L2 trigger corrections applied are presented in Table 1. The results obtained in the four supersamples are shown in Figure 5(a). The measurements from individual supersamples are compatible.

As a systematic check, to measure the size of the systematic biases cancelling due to the quadruple ratio technique, two other quadruple ratios of the eight $u$ spectra were formed. These are the products of four ratios of $u$ spectra of same sign kaons recorded with different setup configurations, therefore any physical asymmetry cancels, while the setup asymmetries do not. The fake slope difference $\Delta g_{LR}$ introduced by global time-variable biases does not cancel if we consider ratios with opposite spectrometer polarities and identical beam line polarities in numerator and denominator, or equivalently, in the adopted notation

$$R_{LR}(u) = (R_{US}(u) \cdot R_{DS}(u))/(R_{UJ}(u) \cdot R_{DJ}(u)). \tag{5}$$

Similarly, the fake slope difference $\Delta g_{UD}$ introduced by the differences of the two beam paths does not cancel if we consider ratios with opposite beam line polarities and identical spectrometer polarities, which look in the following way in the adopted notation:

$$R_{UD}(u) = (R_{US}(u) \cdot R_{UJ}(u))/(R_{DS}(u) \cdot R_{DJ}(u)). \tag{6}$$

The measured fake slope differences $\Delta g_{LR}$ and $\Delta g_{UD}$ in the four supersamples are presented in Figure 4(c) and (d) for both data and MC. The size of these control quantities demonstrates that the cancellation of the first-order systematic biases in (4), due to residual time-variable imperfections in the apparatus, is at the level of a few $10^{-4}$; therefore second order effects are negligible. Moreover, the comparison with MC simulations shows that the asymmetries of the apparatus are understood in terms of local inefficiencies and beam optics variations.

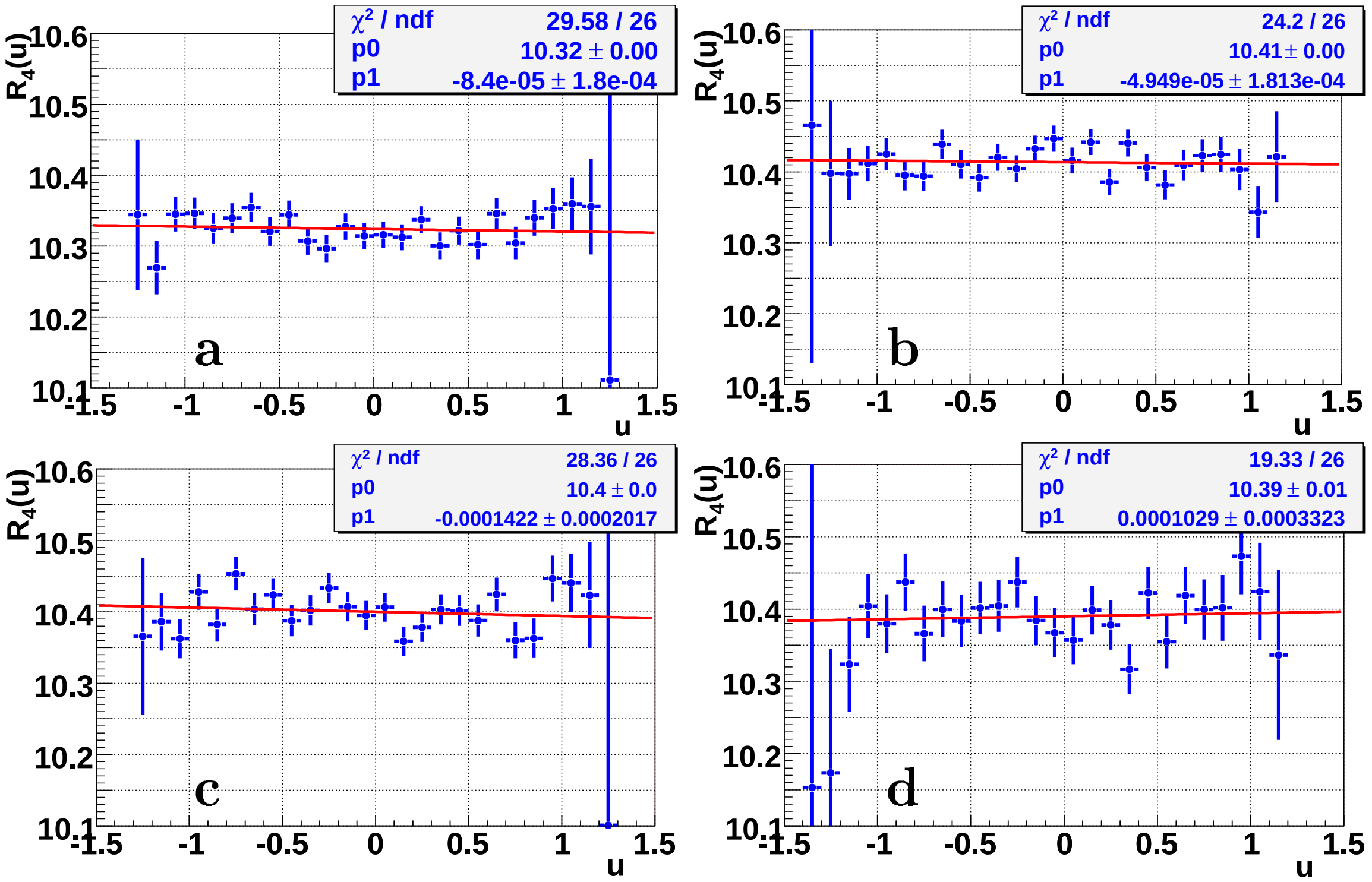

Figure 4: Fitted quadruple ratios $R_4(u)$ for (a) supersample 0; (b) supersample 1; (c) supersample 2; (d) supersample 3. Some points in the extreme bins of $u$ with large errors are outside the scale of the plots.

| Supersample | $K^+ \to \pi^+\pi^+\pi^-$ decays in $10^6$ | $K^- \to \pi^-\pi^-\pi^+$ decays in $10^6$ | $\Delta g \times 10^4$ raw | $\Delta g \times 10^4$ corrected |
|---:|---:|---:|---:|---:|
| 0 | 448.0 | 249.7 | $0.5 \pm 1.4$ | $-0.8 \pm 1.8$ |
| 1 | 270.8 | 150.7 | $-0.4 \pm 1.8$ | $-0.5 \pm 1.8$ |
| 2 | 265.5 | 147.8 | $-1.5 \pm 2.0$ | $-1.4 \pm 2.0$ |
| 3 | 86.1 | 48.0 | $0.2 \pm 3.2$ | $1.0 \pm 3.3$ |
| Total | 1070.4 | 596.2 | $-0.2 \pm 0.9$ | $-0.7 \pm 1.0$ |

Table 1: Statistics selected in each supersample and the measured $\Delta g$: "raw" and "corrected" for trigger efficiency.

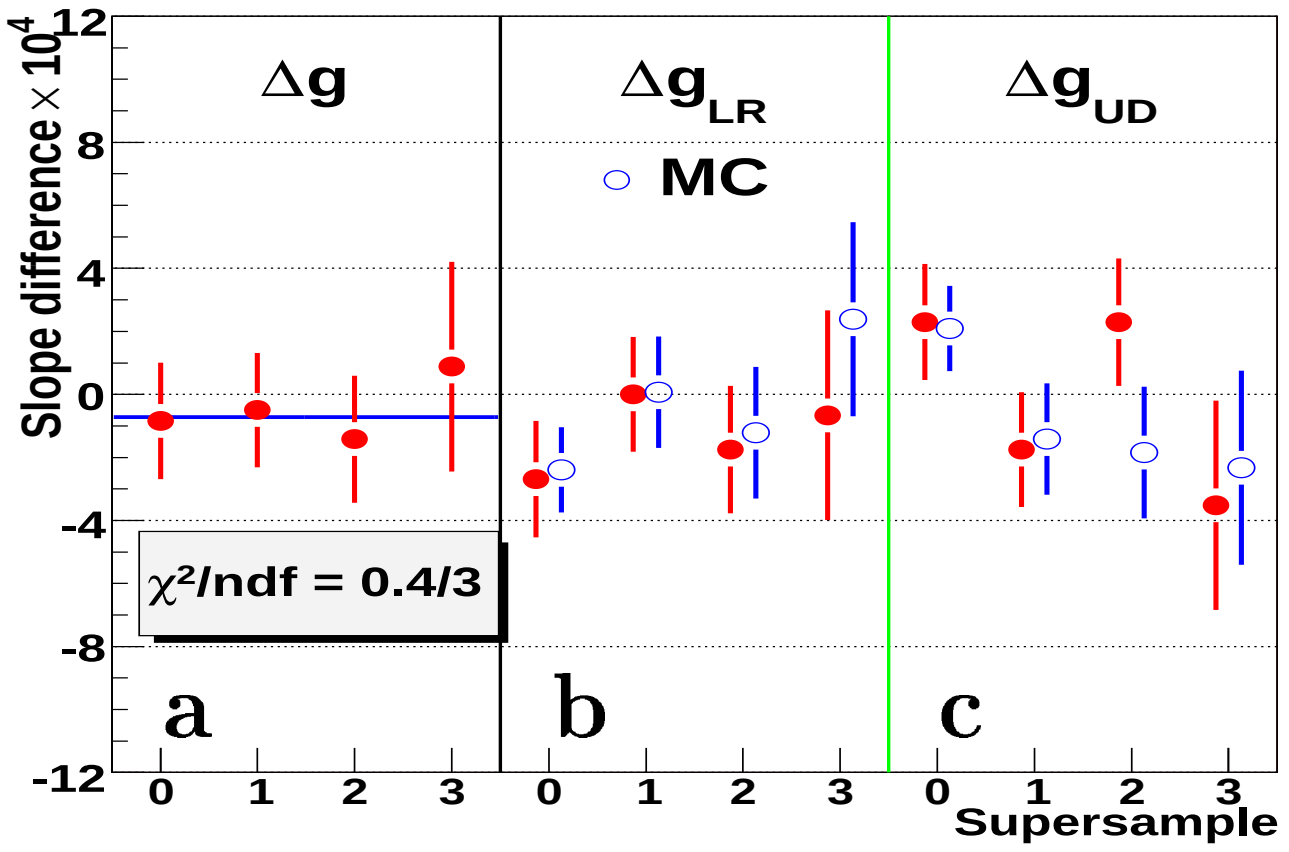

Figure 5: (a) $\Delta g$ measurement in the four supersamples; control quantities (b) $\Delta g_{LR}$ and (c) $\Delta g_{UD}$ corresponding to detector and beam line asymmetries which cancel in quadruple ratio, and their comparison to Monte Carlo.

Another systematic check is given by the slope difference in the $v$ projection of the Dalitz plot which is expected to be at least an order of magnitude smaller than the one in the $u$ projection. The quadruple ratio of $|v|$ spectra similar to (4) was fitted with $f_v(v) = n\cdot(1+4\Delta k\cdot v^2)$ (as a linear term in $v$ is forbidden). A value of $\Delta k = (-0.5\pm0.6_{stat.})\times10^{-4}$ was obtained from this fit, which is compatible with no asymmetry, as expected.

*Limits for residual systematic effects*

The measurement of pion momenta is based on the knowledge of the magnetic field in the spectrometer magnet. The variation of the current in the magnet can be monitored with a relative precision of $5\times10^{-4}$. Smaller variations are continuously controlled with a precision of $\sim 10^{-5}$ by the deviation of the measured charge-averaged kaon mass from the nominal value. A time-dependent correction can be introduced by scaling the reconstructed pion momenta symmetrically for positive and negative tracks. However, the effect is charge symmetric by design, due to the simultaneous beams. A conservative upper limit of $\delta(\Delta g) = 0.1\times10^{-4}$ for the corresponding systematic uncertainty is obtained by comparing results obtained with and without the correction.

In a considerable fraction of the selected events ($\sim 5\%$) at least one of the pions undergoes a $\pi\to\mu\nu$ decay in the decay volume and the spectrometer reconstructs the resulting muon. The tails of the reconstructed $3\pi$ invariant mass distribution are dominated by these events. Rejection of events with $\pi\to\mu\nu$ decays in MC analysis did not lead to any significant change of the result. By varying the accepted $3\pi$ invariant mass interval in a wide range (5–25 MeV/$c^2$) a conservative systematic uncertainty of $\delta(\Delta g) = 0.4\times10^{-4}$ was attributed to effects due to pion decays.

Taking into account that the composition of the beams is not charge symmetric, event distortions caused by pile-up with products of another kaon decay or a beam halo particle traversing the sensitive region of the spectrometer is a potential source of systematic bias. To study the pile-up effects, an accidental activity generator, tuned using the measured kaon fluxes, kinematical distributions and rates of the halo particles, was introduced into the MC and a production of correlated pairs of an initial event and a perturbed event was carried out. No charge-asymmetric effects on $u$ distributions either in the reconstruction or in the second level trigger were observed down to a level of $\delta(\Delta g) = 0.2\times10^{-4}$, limited by MC statistics.

Biases due to resolution effects were studied by using various methods of $u$ variable calculation from the measured track momenta (including a 4C kinematic fit). They differ in the resolution as a function of $u$ itself. The result is stable within $\delta(\Delta g) = 0.2\times10^{-4}$.

Charge-asymmetric material effects have been found to be negligible by evaluating the effect of hadronic interactions in the material in front of and in the chambers taking into account the pion spectra.

A summary of the systematic uncertainties and trigger corrections is presented in Table 2.

| Systematic effect | Correction $\delta(\Delta g) \times 10^4$ | Uncertainty $\delta(\Delta g) \times 10^4$ |
|---|---|---|
| Spectrometer alignment | 0 | $\pm 0.1$ |
| Acceptance and beam geometry | 0 | $\pm 0.3$ |
| Momentum scale | 0 | $\pm 0.1$ |
| Pion decay | 0 | $\pm 0.4$ |
| Pile-up | 0 | $\pm 0.2$ |
| Resolution and fitting | 0 | $\pm 0.2$ |
| Total systematic uncertainty | 0 | $\pm 0.6$ |
| Level 1 trigger | 0 | $\pm 0.4$ |
| Level 2 trigger | $-0.5$ | $\pm 0.5$ |

Table 2: Systematic uncertainties and correction for level 2 trigger inefficiency.

# 5 Conclusions

The difference in the linear slope parameter of the Dalitz plot for $3\pi^\pm$ decays of $K^+$ and $K^-$, measured with the 2003 data sample, is found to be

$$\Delta g = g^+ - g^- = (-0.7 \pm 0.9_{stat.} \pm 0.6_{trig.} \pm 0.6_{syst.}) \times 10^{-4}. \tag{7}$$

Converted to the direct CP violating charge asymmetry in $K^\pm \to 3\pi^\pm$ decays using the PDG value of the Dalitz plot slope $g = -0.2154 \pm 0.0035$ [7], this leads to

$$A_g = (1.7 \pm 2.1_{stat.} \pm 1.4_{trig.} \pm 1.4_{syst.}) \times 10^{-4} = (1.7 \pm 2.9) \times 10^{-4}. \tag{8}$$

The uncertainty due to the trigger inefficiency is of statistical nature; it is dominated by the contribution of supersample 0, when the trigger was still being tuned and the inefficiencies were therefore larger. The precision obtained is limited mainly by the available statistics. The result has more than an order of magnitude better precision than previous measurements and is compatible with the Standard Model predictions. Analysis of the 2004 data will approximately double the data sample and will improve the uncertainties accordingly.

We gratefully acknowledge the CERN SPS accelerator and beam line staff for the excellent performance of the beam. We thank the technical staff of the participating laboratories and universities for their effort in the maintenance and operation of the detectors, and in data processing.